\documentclass[preprint,12pt]{elsarticle}

\usepackage{lineno,hyperref}
\modulolinenumbers[5]
\usepackage{amssymb}
\journal{Physics Letters A}

\begin{document}

\title{Global correlation matrix spectra of the surface temperature of the Oceans from Random Matrix Theory to Poisson fluctuations}

\author[PPGBEA,UFS]{Eucymara F. N. Santos}
\ead{eucymara@gmail.com}

\author[UFRPEDF]{Anderson L. R. Barbosa}
\ead{anderson.barbosa@ufrpe.br}

\author[PPGBEA,UFRPE]{Paulo J. Duarte-Neto\corref{cor1}}
\ead{pjduarteneto@gmail.com}

\address[PPGBEA]{Programa de P\'os-Gradua\c{c}\~ao em Biometria e Estat\'istica Aplicada, Universidade Federal Rural de Pernambuco, 52171-900, Recife-PE, Brazil}
\address[UFS]{Departamento de Estat\'istica e Ci\^encias Atuariais, Universidade Federal de Sergipe, Av. Marechal Rondon, s/n - Jardim Rosa Elze, 49100-000, S\~ao Crist\'ov\~ao - SE}
\address[UFRPEDF]{Departamento de F\'{\i}sica, Universidade Federal Rural de Pernambuco, 52171-900, Recife-PE, Brazil}
\address[UFRPE]{Departamento de Estat\'istica e Inform\'atica, Universidade Federal Rural de Pernambuco, 52171-900, Recife-PE, Brazil}

\cortext[cor1]{Corresponding author}

\date{\today}

\begin{abstract}
In this work we use the random matrix theory (RMT) to correctly describe the behavior of spectral statistical properties of the sea surface temperature of oceans. This oceanographic variable plays an important role in the global climate system. The data were obtained from National Oceanic and Atmospheric Administration (NOAA) and delimited for the period 1982 to 2016. The results show that oceanographic systems presented specific $\beta$ values that can be used to classify each ocean according to its correlation behavior. The nearest-neighbors spacing of correlation matrix for north, central and south of the three oceans get close to a  RMT distribution. However, the regions delimited in the Antarctic pole exhibited the distribution of the nearest-neighbors spacing well described by the Poisson model, which shows a {\it statistical change} of RMT to Poisson fluctuations. 
\end{abstract}


\maketitle


\section{Introduction}

{\it The Ocean and its interaction with the atmosphere are significant components that influence Earth's weather and climate \cite{Biggetal}. The presence of a pattern in ocean variables as sea surface temperature (SST), sea level pressure (SLP) and wind speed (WS) allows to understand these climatic variations more clearly. For instance, the SST presents a chaotic dynamic \cite{Swarnali} and is fundamental for the understanding of the behavior of a meteorological system, important in several applications in the maritime area. Cyclones, hurricanes, rainfall, ocean currents patterns, weather forecasts, surface energy flux, atmospheric ocean interactions are characteristics studied and influenced by SST. Thus it plays an important role in the study of the variability of patterns in the global climate system \cite{LI201314,NASA}.}

Among the many characteristics of oceanographic variables, their empirical correlation matrices are object of study in multivariate time series analysis in atmospheric sciences and many other fields. Their statistical properties are capable to separate the signal from ‘‘noise’’. For instance, the data matrices are often subject of empirical orthogonal function (EOF) method (frequently called as Principal Component Analysis), and they are used to identify and study the different modes of variability or correlation of atmospheric variables in large scale patterns \cite{Preisendorfer}. In this case, the dominant mode associated to a geographical location would correspond to the EOF with the largest eigenvalue.

Evaluating the density distribution of SLP, SST and WS eingenvalues, Santhanam and Patra \cite{PhysRevE.64.016102} showed that the empirical correlation matrices that arise in atmospheric sciences can be modeled as a random matrix chosen from an appropriate ensemble. They found that the Gaussian orthogonal ensemble (GOE) is appropriate for the mean SLP and SST correlations and the Gaussian unitary ensemble (GUE) is appropriate for pseudo-wind-stress vectors. That is, these variables present good agreement with the universality classes of the Random Matrix Theory (RMT) \cite{Mehta,RevModPhys.82.2845}. Therefore, this technique has great potential to be applied in separating the random modes from the physically significant modes of the correlation matrix. RMT  was first used in physics by Eugene Wigner in the 1950s in nuclear physics, with the objective of studying the statistical properties of heavy atom spectra \cite{10.2307/1970079}. It has been successfully applied in a variety of multivariate datasets in the various areas of knowledge, such as finance \cite{PhysRevLett.83.1471,PLEROU2000374,PLEROU2001175,PhysRevE.65.066126,STOSIC2018499}, human health \cite{PhysRevLett.91.198104}, biological systems \cite{LUO2006420}, protein dynamics \cite{PhysRevLett.103.268101,AGRAWAL2014359,PhysRevE.76.026109}, subway system \cite{PhysRevE.96.030101}, HIV infection \cite{GONZALEZ20172912} and seismic vibrations \cite{CHATTERJEE20181352}. 

Besides the capability to separate the random modes, RMT is used as a diagnostic tool to find the presence
of transitions in the spectral distribution behavior. Srivastava et. al. \cite{PhysRevLett.116.054101} show that the spectral fluctuations of non-interacting chaotic systems are well described by the Poisson distribution while when the interaction is introduced the fluctuations are well described by the RMT. The similar behavior manifest in fluctuation of Bombay Stock Exchange index data \cite{CHATTERJEE2019122189}. On the other hand,  in all the cases studied by \cite{PhysRevE.64.016102}, the empirical histograms did not follow the Poisson curves at all. For this reason, these authors questioned whether it would be possible to observe limits of the spectra of correlation matrices of the ocean SST that can exhibit Poisson spacing, behavior that has not been yet observed.

Following this questioning, the technique of RMT was applied here to analyze the correlation structure of the matrices of SST data series derived from a more comprehensive area of Pacific, Atlantic and Indian Oceans, considering north, central and south regions. Our wider area allowed for proper evaluations of the different correlation behavior of SST from global Oceans, and of the presence of {\it statistical change of RMT-to-Poisson} in the studied area. 
\begin{figure}
\centering
 \includegraphics[width=1.0\linewidth]{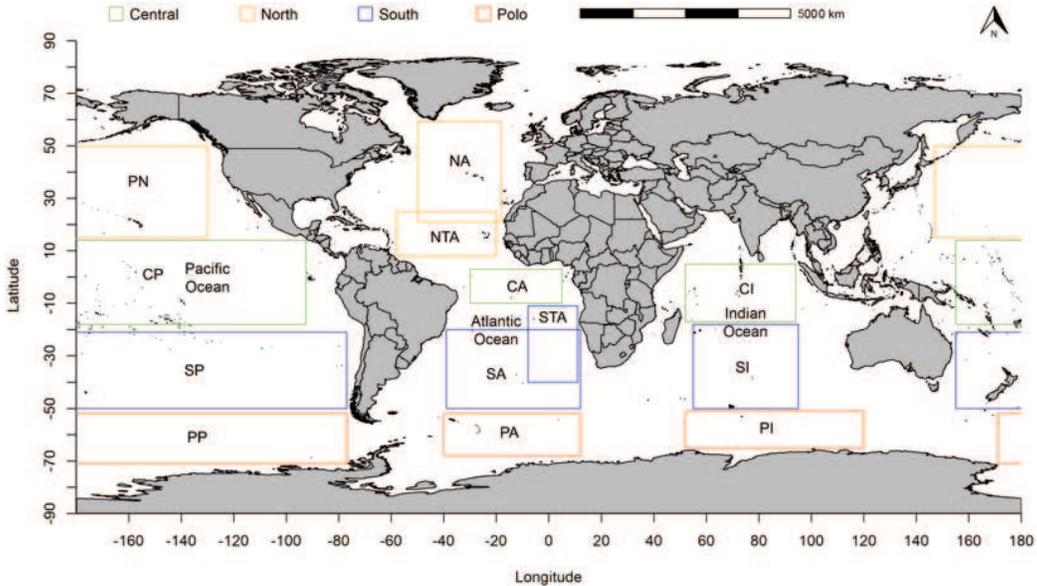}
\caption{ World map with the matrix boundaries of NOAA data for the Pacific, Atlantic and Indian Oceans. Each oceans region can be understudy as a matrix and the labels are described in Table \ref{tabela}.} \label{fig1}
\end{figure}

In the following sections, the concepts and properties of the RMT based on the behavior of the distribution of eigenvalue spacing in several bounded regions in the Pacific, Atlantic and Indian Oceans, are presented. It was demonstrated that all oceanographic systems presented a specific eigenvalue spacing distribution that can be used to classify each ocean according to its correlation behavior. Besides, we show that north, central and south ocean regions exhibit an eigenvalue spacing distribution approaching a RMT, while pole ocean regions approaches a Poisson distribution. This represents a {\it statistical change} in behavior not yet observed for this dynamical system and answers the issue raised by Santhanam and Patra \cite{PhysRevE.64.016102}.  Therefore, the results indicate that the general dynamic of SST can be understood as a set weakly interacting dynamic subsystems with Poisson fluctuations or one with strong interactions, with RMT fluctuations \cite{PhysRevLett.116.054101,CHATTERJEE2019122189}.

\section{Ocean Surface Temperature Data Arrays}

The SST data was obtained from the National Oceanic and Atmospheric Administration (NOAA) website \cite{NOAA}, where various oceanographic measurements have been systematically provided since 1981. In the NOAA website, a global and continuous coverage map is presented with daily temporal and spatial resolution. The data are obtained by combining observations from different platforms (satellites, ships and buoys), corrected through the optimum interpolation algorithm for large-scale satellite bias adjustment in relation to in-situ moored and drifting data, and also to fill space gaps \cite{doi:10.1175/2007JCLI1824.1}.

The complete NOAA SST data matrix has spacing of $0.25^0$ between geographic coordinates, with dimensions of 720 latitudes (rows) with grids of $89.875^0$ N to $89.875^0$ S, and 1440 longitudes (columns) with grids of $0.125^0$ E at $379.5^0$ W, as illustrated in the Fig. (\ref{fig1}). 

It is well known that the north, central, south and pole regions of the Pacific, Atlantic and Indian Oceans have different climate dynamics \cite{Preisendorfer}. Hence, it is convenient to divide the oceans in regions as showing in the Fig. (\ref{fig1}). Each region can be understudy as a matrix with dimension conveniently selected as showing in Table \ref{tabela} and \ref{tabela2}. 

The ocean data for the period 1982 to 2016 was selected, totaling 12,784 days, so that each region of Table \ref{tabela} corresponds to an ensemble of 12784 matrices. These matrices were used to obtain the correlation matrix, which is then analyzed within the RMT framework, as described in the following section.

\begin{figure}
\centering
    \includegraphics[width=0.49\linewidth]{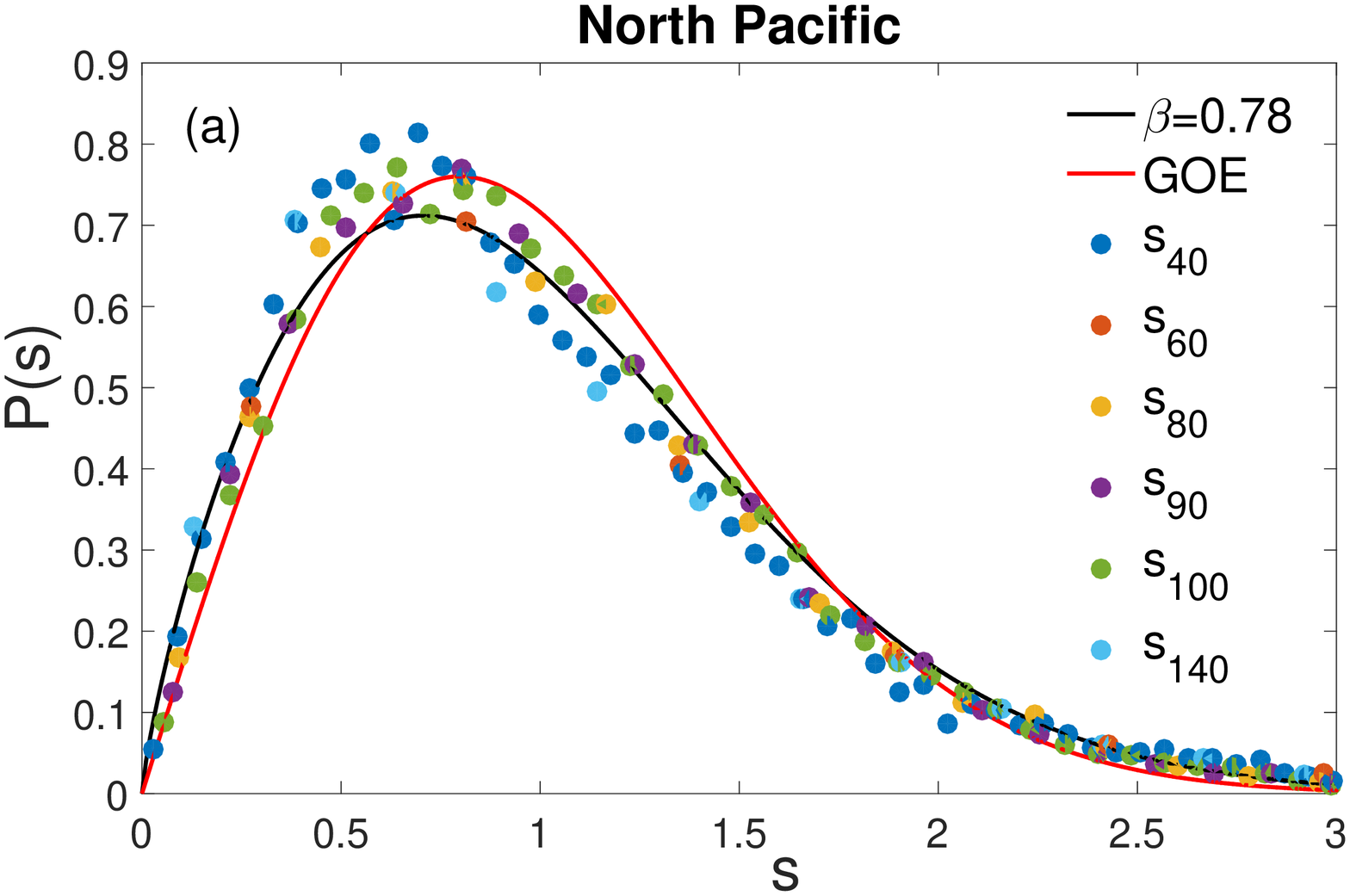}
    \includegraphics[width=0.49\linewidth]{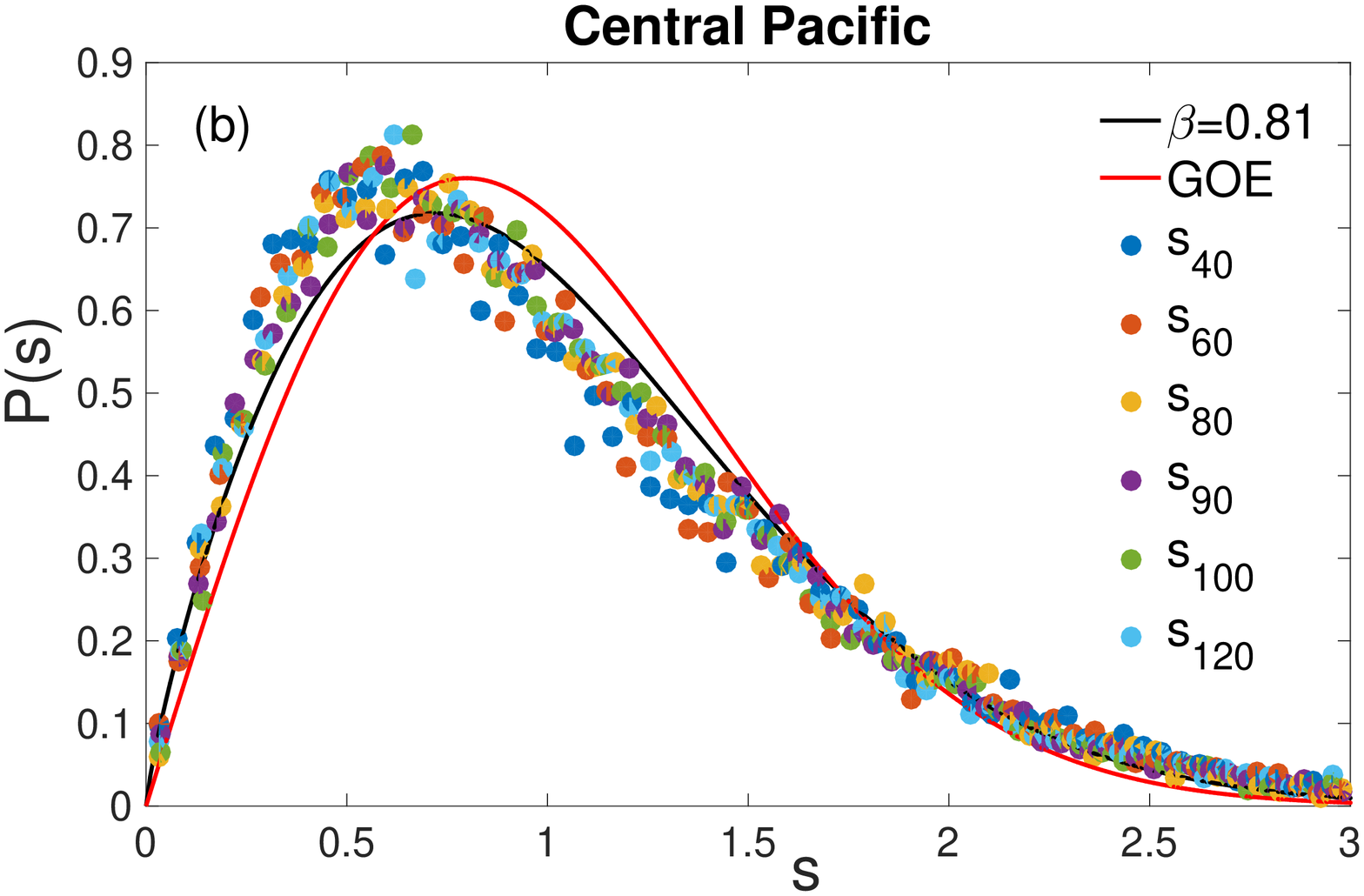}
    \includegraphics[width=0.49\linewidth]{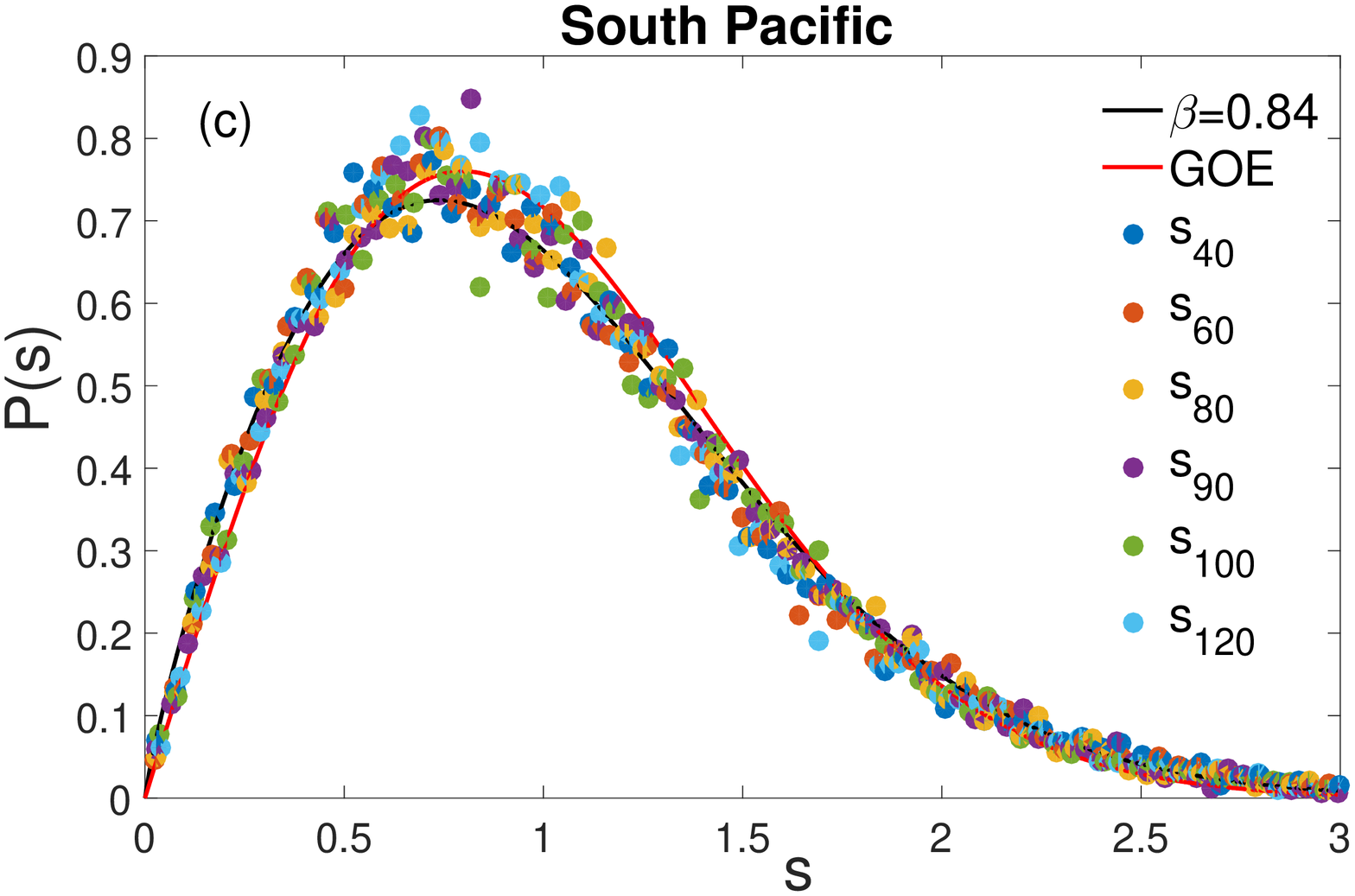}
    \includegraphics[width=0.49\linewidth]{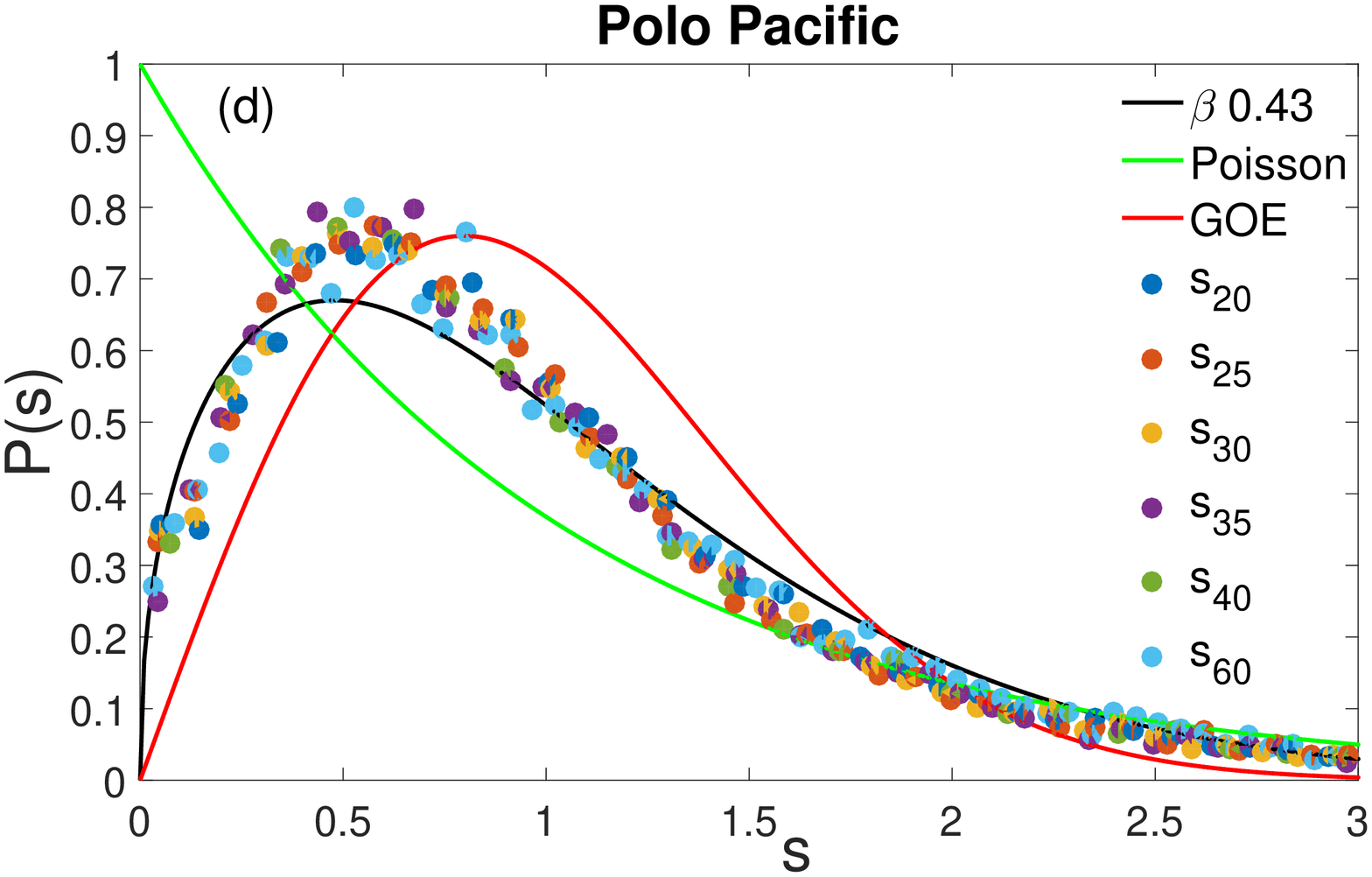}
\caption{ The figures (a - d) show the nearest-neighbor spacing histogram $s_j$ for each oceanic region of Pacific Ocean. The black line is obtained from Brody equation (\ref{P}) while the red and green lines are GOE distribution (\ref{GOE}) and Poisson distribution (\ref{Poi}), respectively. Note that the figures (a), (b) and (c) get close to GOE while (d) tends to Poisson.} \label{fig2}
\end{figure}

\begin{table}
\centering
\caption{Geographic coordinates of the data matrices for each of the studied oceans obtained from the National Oceanic and Atmospheric Administration (NOAA). The label is used in the Fig. (\ref{fig1}).}\label{tabela}

\begin{tabular}{c|c|c|c}
\hline
    \hline
Label & Oceans & Latitude  & Longitude   \\\hline
    \hline
NP & North Pacific & ($50.875^0$ N; $15.125^0$ N) & ($229.875^0$ W; $145.375^0$ E)  \\
NA & North Atlantic & ($59.625^0$ N; $20.125^0$ N) & ($307.375^0$ W; $342.875^0$ W)  \\
NTA & North Tropical Atlantic & ($25.875^0$ N; $8.625^0$ N) & ($299.625^0$ W; $325.125^0$ W)  \\\hline
CP & Central Pacific & ($15.625^0$ N; $15.875^0$ S) & ($266.625^0$ W; $152.375^0$ E)\\
CA & Central Atlantic & ($4.125^0$ N; $10.875^0$ N) & ($325.375^0$ W; $8.875^0$ E) \\
CI & Central Indian & ($5.875^0$ N; $15.875^0$ S) & ($50.625^0$ E; $95.125^0$ E)  \\\hline
SP & South Pacific & ($50.875^0$ S; $20.125^0$ S) & ($284.375^0$ W; $153.625^0$ E) \\
SA & South Atlantic & ($11.375^0$ S; $40.125^0$ S) & ($342.875^0$ W; $11.625^0$ E)  \\
STA & South Tropical Atlantic & ($20.125^0$ S; $50.875^0$ S) & ($319.875^0$ W; $13.125^0$ E)  \\
SI & South Indian & ($15.875^0$ S; $50.875^0$ S) & ($50.625^0$ E; $94.875^0$ E) \\\hline
PP & Pole Pacific & ($51.125^0$ S; $71.625^0$ S) & ($284.625^0$ W; $170.875^0$ E) \\
PA & Pole Atlantic & ($51.125^0$ S; $69.875^0$ S) & ($319.875^0$ W; $13.125^0$ E)  \\
PI & Pole Indian & ($50.125^0$ S; $65.625^0$ S) & ($49.875^0$ E; $120.125^0$ E) \\\hline   
\hline

\end{tabular}

\end{table}

\begin{table}
\centering
\caption{Dimension of data matrices for each of the studied oceans obtained from the National Oceanic and Atmospheric Administration (NOAA). The label is used in the Fig. (\ref{fig1}).}\label{tabela2}

\begin{tabular}{c|c|c}
\hline
    \hline
Label & Oceans & Matrix Dimension ($M\times T$) \\\hline
    \hline
NP & North Pacific  & $144\times 399$ \\
NA & North Atlantic  & $159\times143$ \\
NTA & North Tropical Atlantic  & $116\times116$ \\\hline
CP & Central Pacific  & $127\times458$\\
CA & Central Atlantic  & $61\times175$\\
CI & Central Indian  & $88\times179$ \\\hline
SP & South Pacific  & $124\times524$ \\
SA & South Atlantic  & $124\times214$ \\
STA & South Tropical Atlantic & $70\times123$ \\
SI & South Indian & $141\times178$ \\\hline
PP & Pole Pacific  & $83\times456$ \\
PA & Pole Atlantic  & $76\times214$ \\
PI & Pole Indian  & $63\times282$ \\\hline   
\hline

\end{tabular}
\end{table}

\begin{figure}
\centering
    \includegraphics[width=0.49\linewidth]{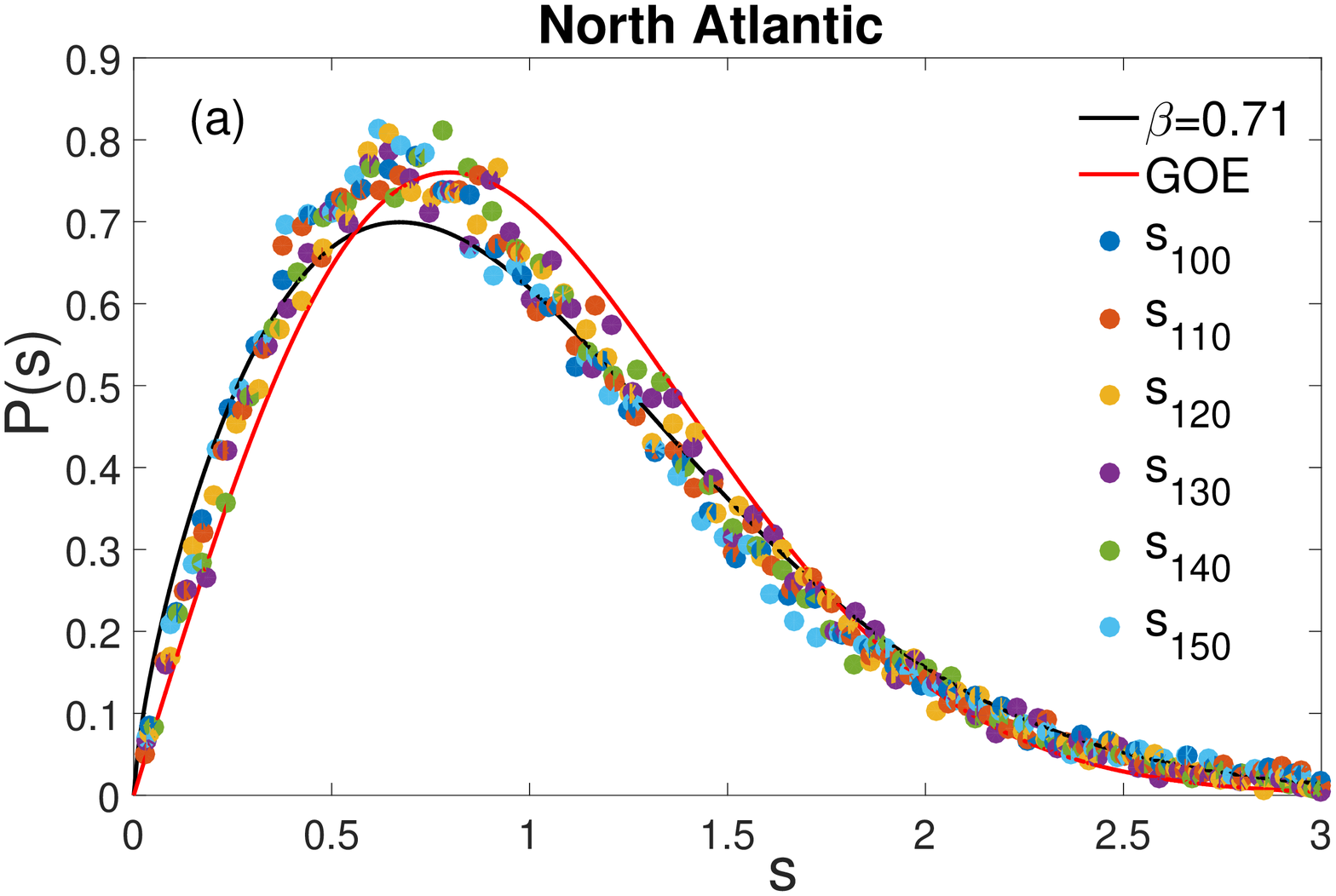}
    \includegraphics[width=0.49\linewidth]{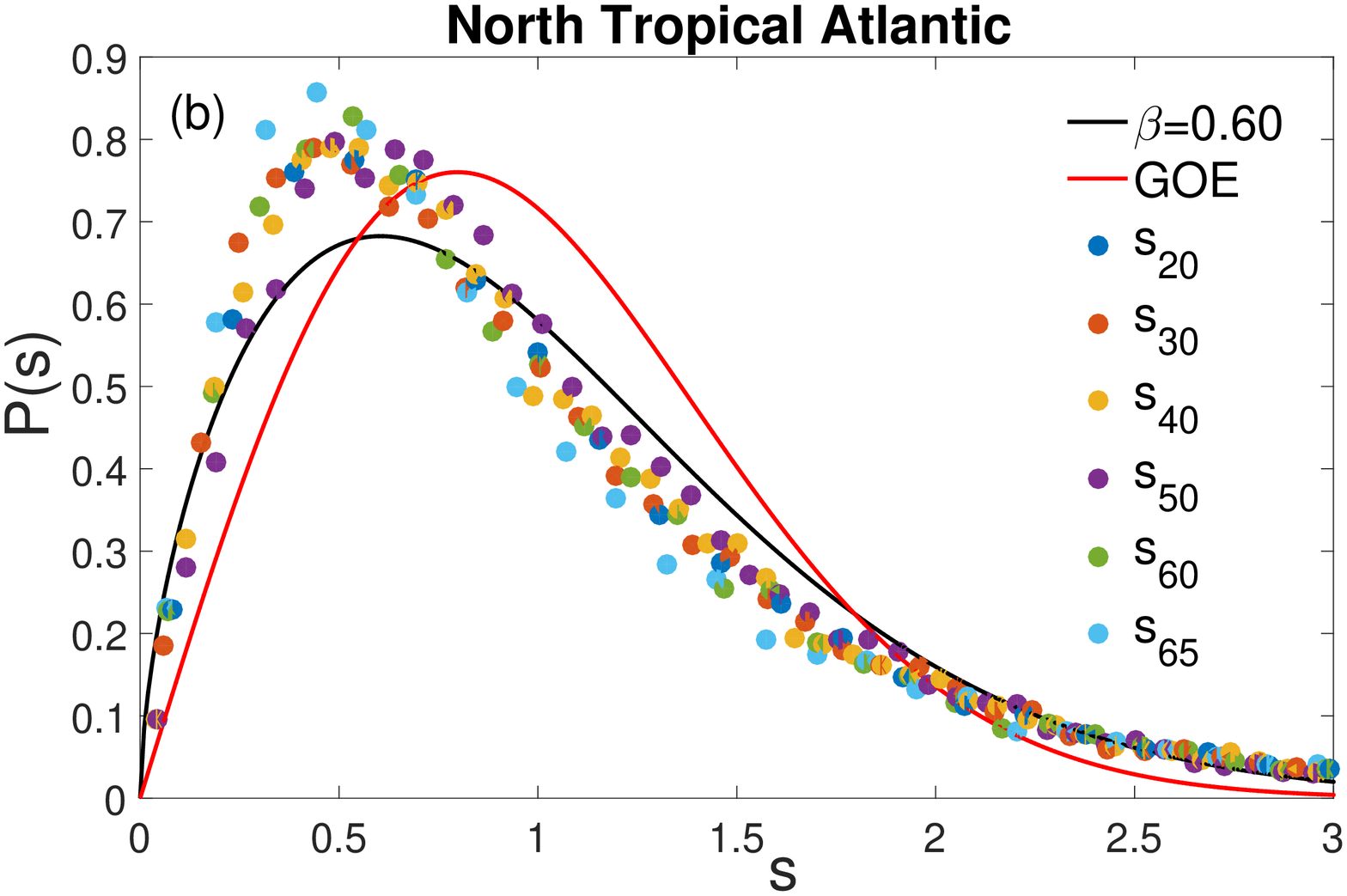}
    \includegraphics[width=0.49\linewidth]{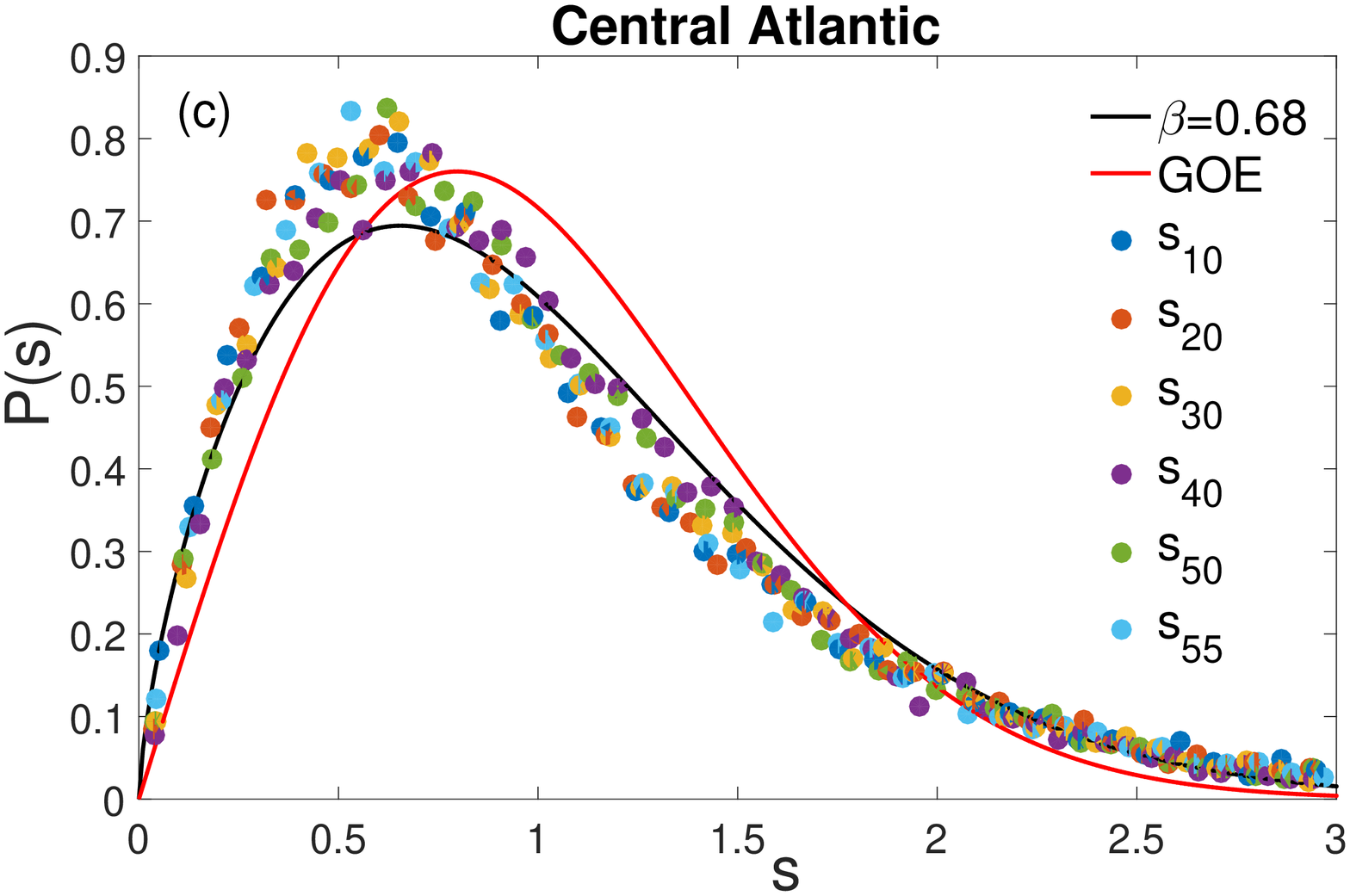}
    \includegraphics[width=0.49\linewidth]{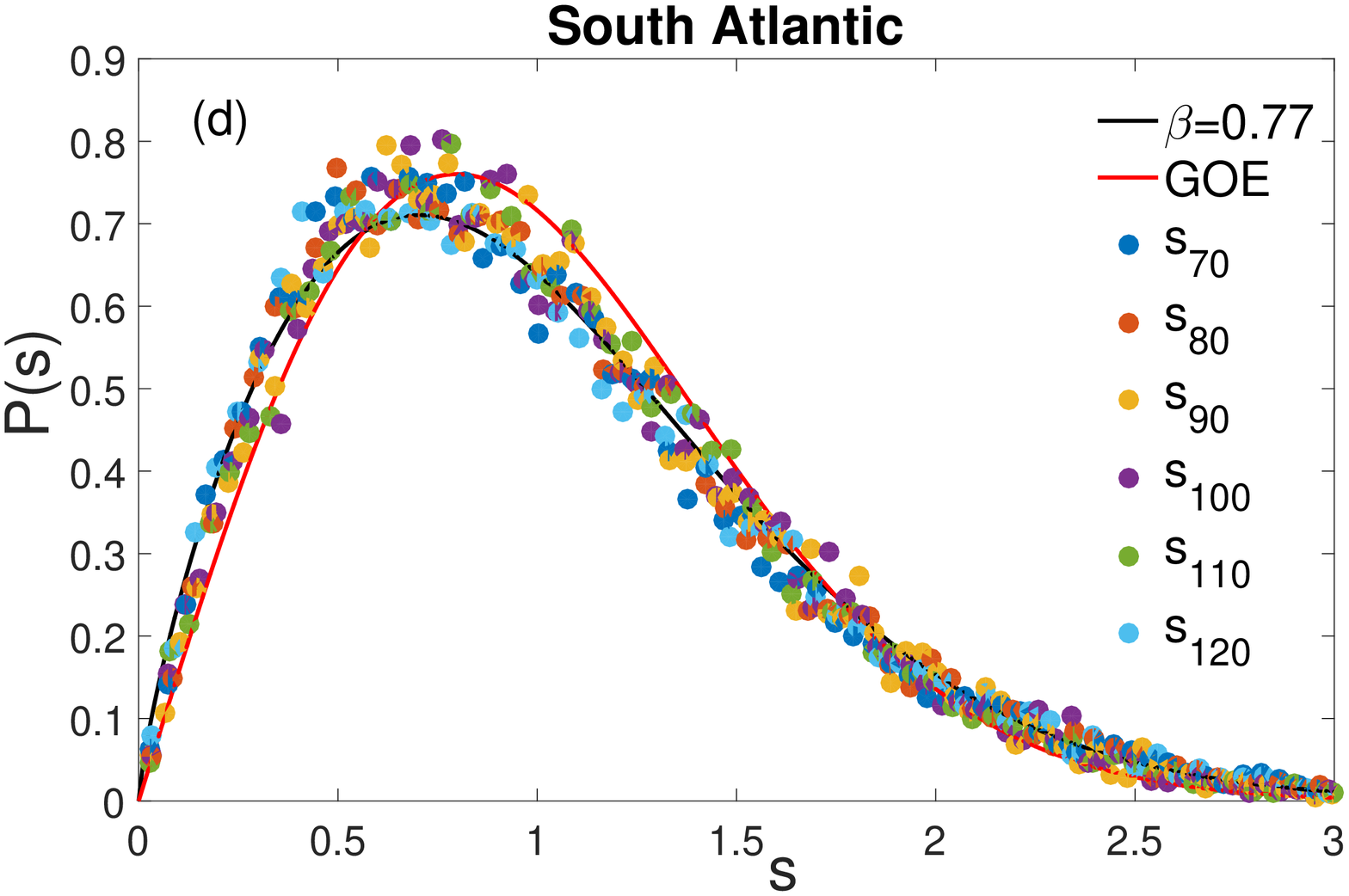}
    \includegraphics[width=0.49\linewidth]{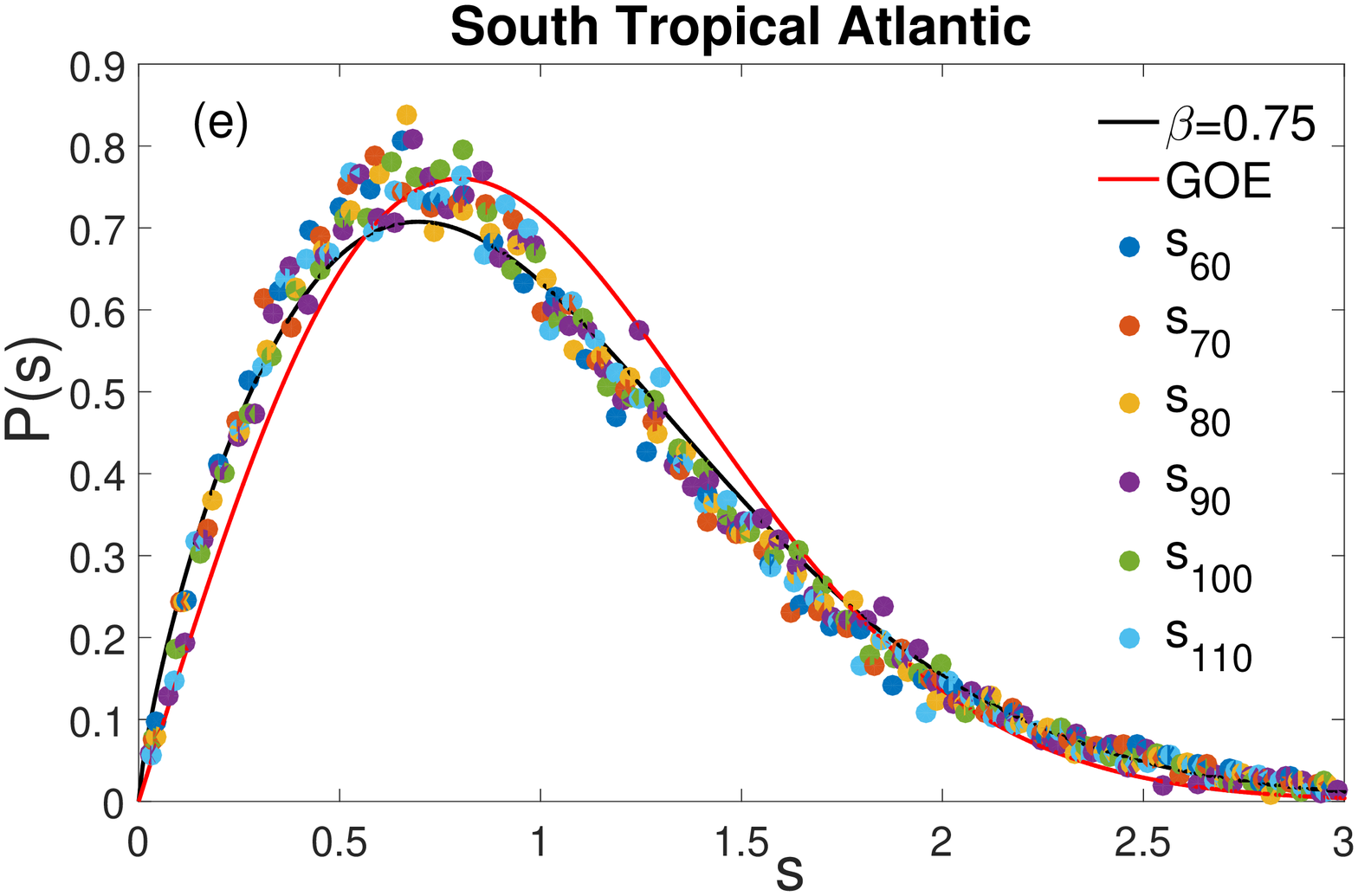}
    \includegraphics[width=0.49\linewidth]{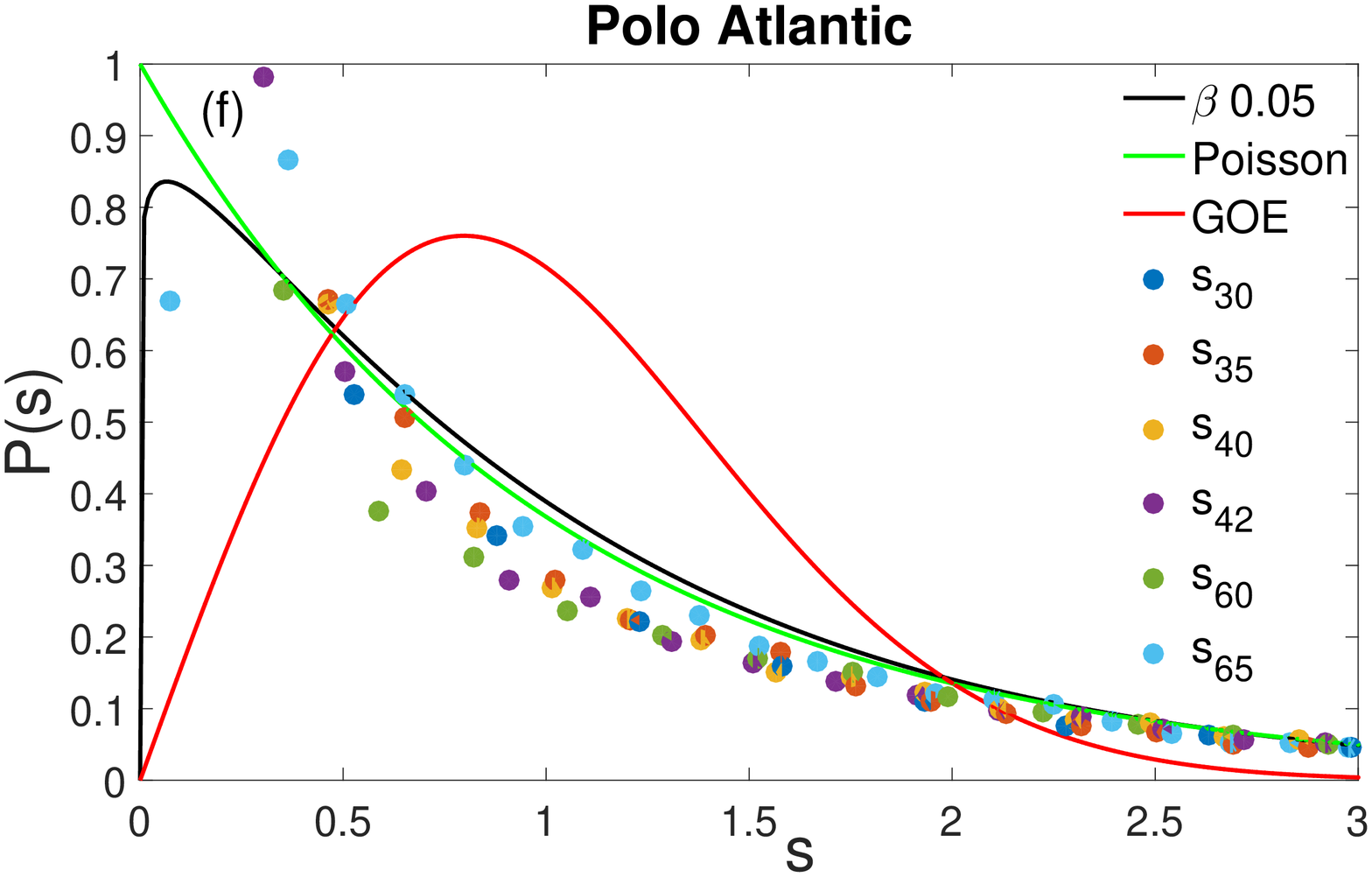}
\caption{ The figures (a - f) show the nearest-neighbor spacing histogram $s_j$ for each oceanic region of Atlantic Ocean. The black line is obtained from Brody equation (\ref{P}) while the red and green lines are GOE distribution (\ref{GOE}) and Poisson distribution (\ref{Poi}), respectively. Note that the figures (a - e) get close to GOE while (f) tends to Poisson.} \label{fig3}
\end{figure}
\section{Random Matrix Theory}

The description of the systems can be defined by an ensemble of random matrices. 
The realization is done through the statistics of the eigenvalue densities fluctuations obtained from the diagonalization of the matrices, whose elements are randomly distributed \cite{RevModPhys.82.2845}. With the aim to study the SST time dynamic of oceans, we introduce here a statistical model to analyze the eigenvalue spacing spectra of correlation matrix obtained from SST ensemble. 

Let $X$ be a matrix with dimension $M \times T$, where $M$ is the number of rows (latitudes), and $T$ is the number of columns (longitude), see Table \ref{tabela2}. From the SST data, we can build an ensemble with 12,784 matrices, one matrix for each day during the years 1982 to 2016. Hence, we can calculate the correlation matrix $C$  \cite{PhysRevLett.103.268101} given by
 \begin{equation}
     C=\frac{XX^t}{T} \label{C},
\end{equation}
where $t$ denotes the matrix transpose. 
From Eq. (\ref{C}), one can compute the eigenvalues $\lambda_i$, which are ordered $\lambda_1 \leq \lambda_2 \leq \dots \leq \lambda_M$ by size. 

From eigenvalues of the correlation matrices, one can obtained the nearest-neighbor spacing given by  
 \begin{equation}
     s_i = \frac{\lambda_{i+1}-\lambda_i}{ \langle \lambda_{i+1}-\lambda_i \rangle},i=1,\dots,M \label{s}
\end{equation}
where $ \langle \lambda_{i+1}-\lambda_i \rangle$ denotes the average of the 12,784 consecutive eigenvalue pair differences. The nearest-neighbor spacing distribution can well described by  Brody distribution \cite{Brody1973}. 

The Brody distribution has been used to describe the energy levels of the nearest-neighbor statistic, represented by the parameter $\beta$, which permits to classify the correlation of the system in relation to its probability distribution, and also describes a direct transition from Poisson to GOE behavior \cite{PhysRevE.72.045204}. The Brody distribution probability density function is given by:
\begin{equation}
     P(s_i)=c(\beta)(1+\beta)s_i^\beta e^{-c(\beta) s_i^{1+\beta}}\label{P}
\end{equation}  
where 
$$c(\beta) = \left[\Gamma\left(\frac{2+\beta}{1+\beta}\right)\right]^{\beta+1}$$ 
and $\Gamma(x)$ is the Gamma function.
The parameter $\beta$ is a measure of the repulsion intensity between neighboring levels, varying from 0 to 1. When level repulsion is not present, there is no correlation between them, that is, $\beta \rightarrow 0$. However, when level repulsion is present there are correlated, hence $\beta  \rightarrow 1$. The latter means that they exhibit more randomness in the structure of the SST matrices.

For $\beta = 0$ Eq. (\ref{P}) reduces the Poisson distribution
\begin{equation}
     P(s_i)= e^{-s_i}.\label{Poi}
\end{equation} 
However, if $\beta = 1$, the system is represented by  the Gaussian Orthogonal Ensemble (GOE), which represents a universal class described by Wigner for matrices whose elements have real inputs. This means that the maximum correlation exists in the structure of the set of time series. The GOE is given by 
\begin{equation}
     P(s_i)=\frac{\pi}{2}s_i e^{-\frac{\pi}{4} s_i^{2}}. \label{GOE}
\end{equation}  

\begin{figure}
\centering
    \includegraphics[width=0.49\linewidth]{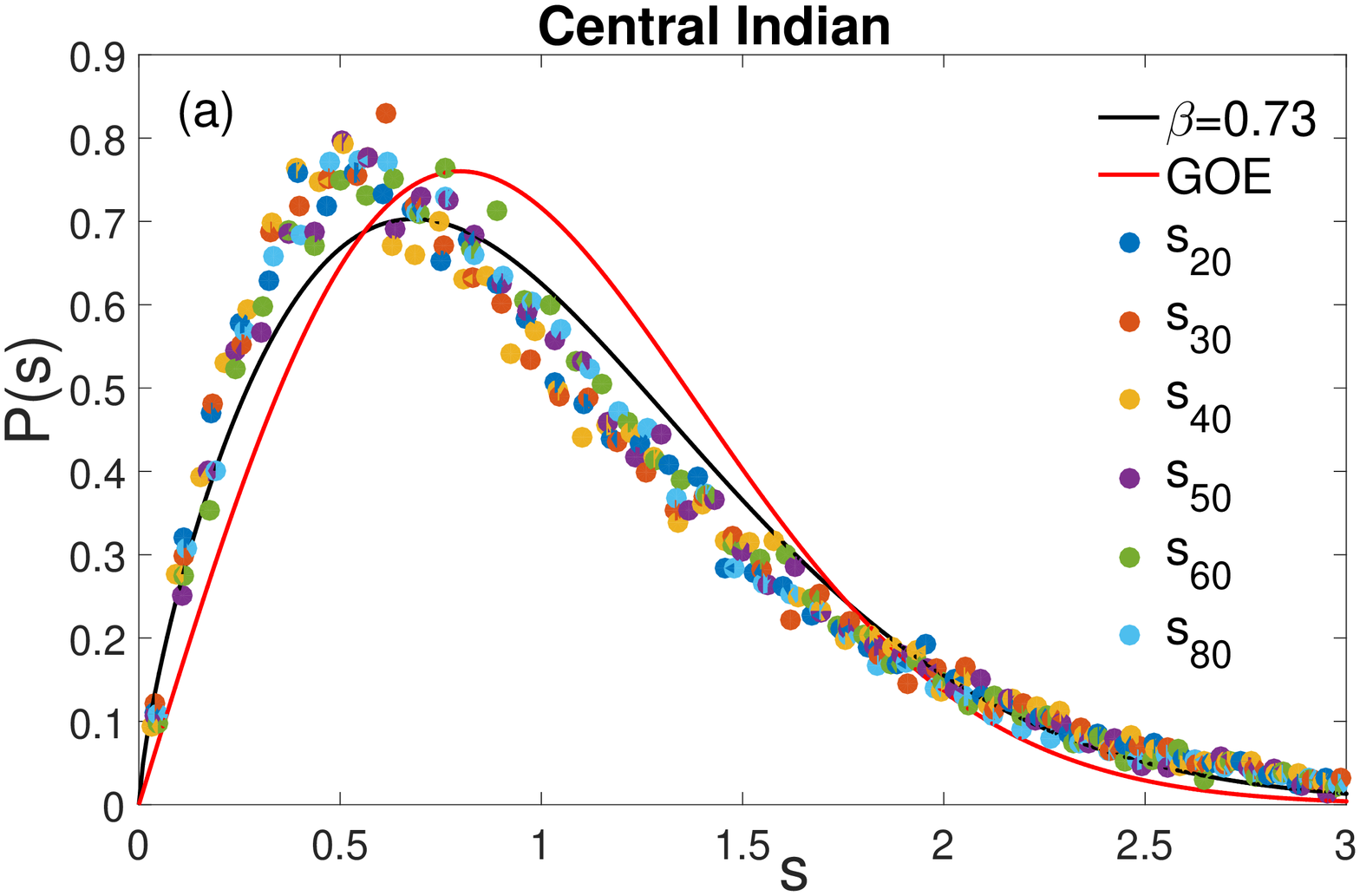}
    \includegraphics[width=0.49\linewidth]{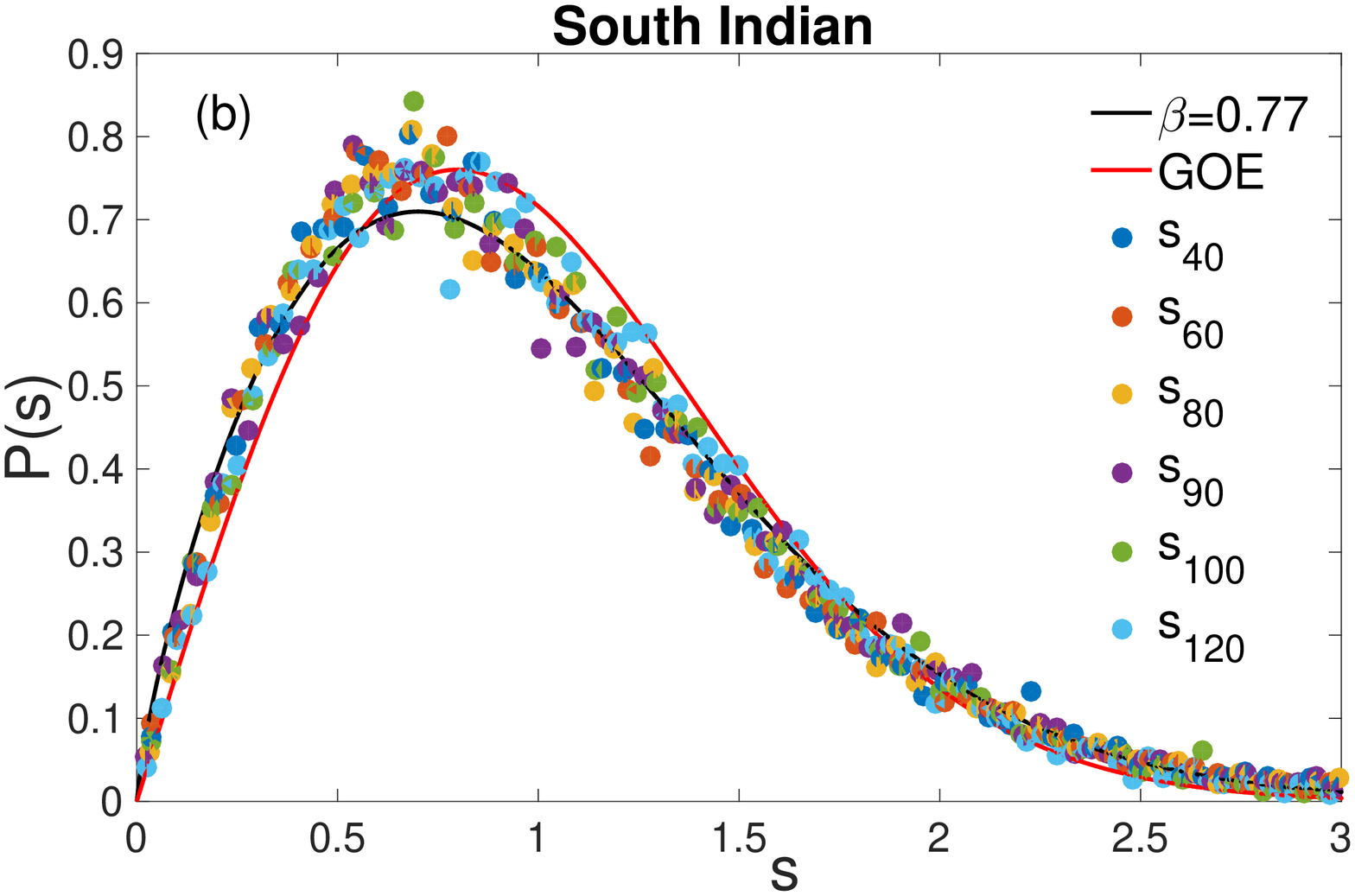}
    \includegraphics[width=0.49\linewidth]{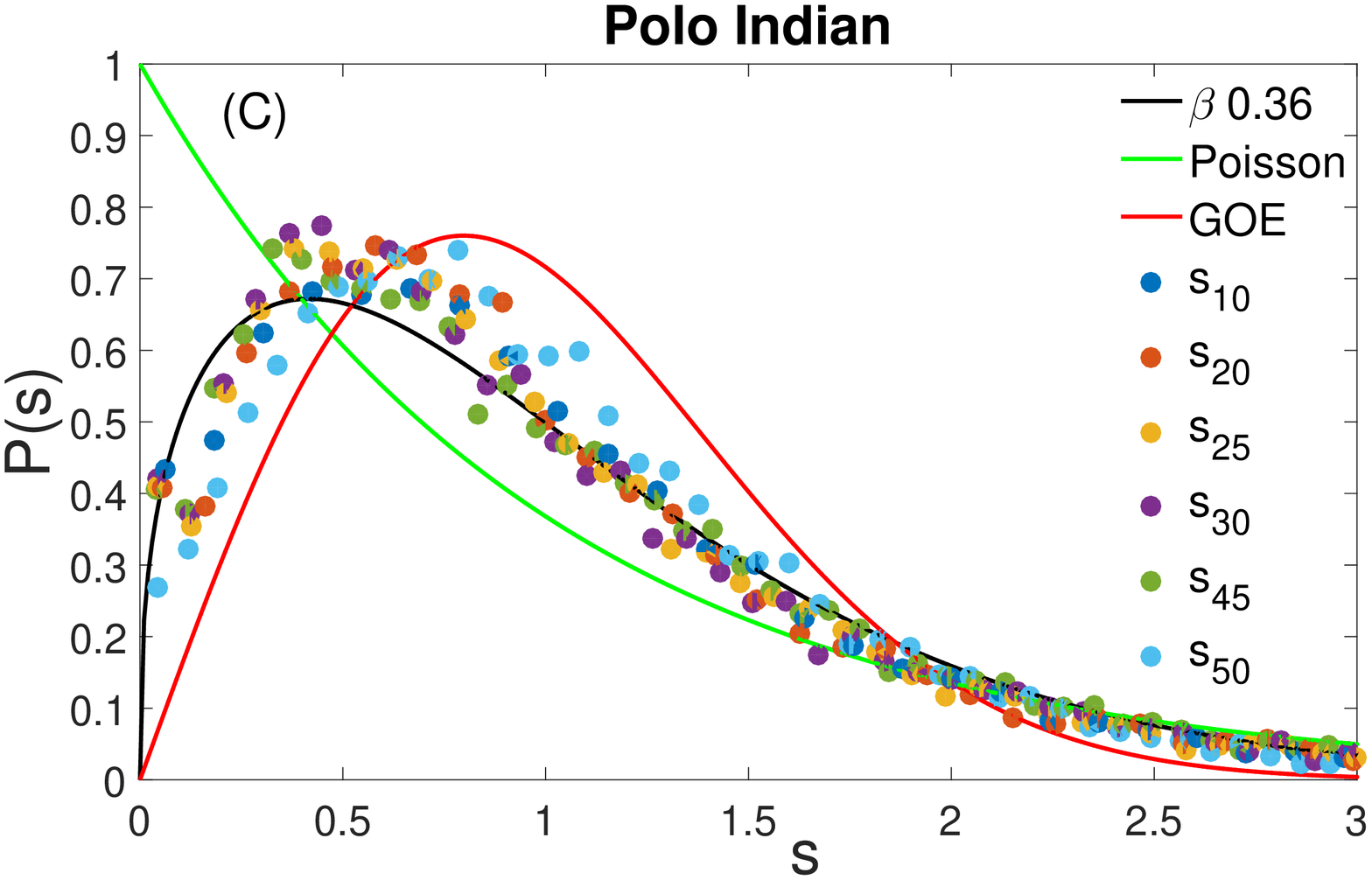}
\caption{ The figures (a - c) show the nearest-neighbor spacing histogram $s_j$ for each oceanic region of Indian Ocean. The black line is obtained from Brody equation (\ref{P}) while the red and green lines are GOE distribution (\ref{GOE}) and Poisson distribution (\ref{Poi}), respectively. Note that the figures (a) and (b) get close to GOE while (c) tends to Poisson.} \label{fig4}
\end{figure}

\section{Results}

The technique of RMT was used to analyze the general behavior of the distribution of eigenvalue spacing in the north, central, south and pole regions of the oceans during the 35 years of observation, summing up 12,784 arrays. Fig. (\ref{fig2}), (\ref{fig3}) and (\ref{fig4}) show the spacing distribution between the nearest-neighbor eigenvalues for Pacific, Atlantic and Indian Oceans, respectively. The number of points depends on the size of each set of matrices and, in turn, the dimensions of the matrices depend on the region and the ocean (Tables \ref{tabela} and  \ref{tabela2}).

The Fig. (\ref{fig2}) shows the histograms of nearest-neighbor eigenvalues for Pacific Ocean. We used Eq. (\ref{P}) to adjust the best fit (black line) of data, besides, the GOE (red line) and Poisson (green line) are plotted for comparison. {\it To obtain the best adjustment, we have performed fits of all $s_i$ histograms ($i=1,\dots,M$) and calculated the average over $\beta_i$, $\beta=\sum_{i=1}^M \beta_i/M$. That means that the best fit (black line) is the  Eq. (\ref{P}) with $\beta$ average. }  

The values found for parameter $\beta$ for the Pacific Oceans were $\beta = 0.79\pm0.01$ (north, Fig.(\ref{fig2}-a)), $\beta = 0.81\pm0.07$ (central, Fig. (\ref{fig2}-b)), $\beta = 0.84\pm0.01$ (south, Fig. (\ref{fig2}-c)), and $\beta = 0.43\pm0.04$ (pole  Fig. (\ref{fig2}-d)). {\it The first three values of $\beta$ are rather close each other, which means that these regions of Pacific Ocean have similar SST dynamics. Furthermore, correlation matrix eigenvalues have a strong repulsion which makes eigenvalue distributions so close to the GOE. The last value of $\beta$ for Pacific Oceans at the pole is significantly lower than the others, which indicates that SST dynamics of the pole region deserves more attention. The value $\beta = 0.43$ means that correlation matrix eigenvalues have a slight repulsion as opposed to other regions. However, this slight repulsion is sufficiently for the eigenvalue distributions to demonstrate a subtle deviation from the Poisson distribution. Hence, the pole region can be understood as set of dynamic subsystems with a slight interaction, while the other three regions can be attributed strong interaction.    }

The Atlantic Ocean was delimited in several matrices due to the narrow oceanographic area in the central region. Besides, the intertropical zone is an important meteorological system that operates in the tropics and, therefore, it was represented by 3 matrices: tropical north, central and tropical south (Table \ref{tabela}; Fig. \ref{fig1}). This division accommodated both the shape of the central region and its climate diversification.
The best fit and proximity to the GOE distribution are found for the delimitation of the south ($\beta = 0.77\pm0.01$, Fig. (\ref{fig3}-d)), as the south Pacific. The $\beta$ values for the others Atlantic systems were: $\beta = 0.71\pm0.01$ (north, Fig. (\ref{fig3}-a)), $\beta = 0.60\pm0.03$ (tropical north, Fig. (\ref{fig3}-b)), $\beta = 0.68\pm0.03$ (central, Fig. (\ref{fig3}-c)) $\beta = 0.75\pm0.01$ (tropical south, Fig. (\ref{fig3}-e)) and $\beta = 0.05\pm0.003$ (pole, Fig. (\ref{fig3}-f)). Comparing their $\beta$ values, one can verify that the tropical north is closer to the central region, while the tropical south is closer to the south region, possibly because the southern tropical region presents a greater superposition with the south. In contrast, the Atlantic pole spacing adjustments were very close to the green curve of the theoretical Poisson distribution Fig. (\ref{fig3}-f).  This indicates that the pole can be understood as a set of dynamic subsystems with weak interaction, which means that what happens in one part of pole does not propagate to another \cite{PhysRevLett.116.054101}.

Among the Indian systems, the south region showed the best approximation to the GOE curve ($\beta = 0.77\pm0.02$, Fig. (\ref{fig4}-b)), as also observed for the Pacific and Atlantic Oceans. The other regions presented the following results: $\beta = 0.73\pm0.06$ (central, Fig. (\ref{fig4}-a)), and $\beta = 0.36\pm0.05$ (pole, Fig. (\ref{fig4}-c)). {\it The central region departs somewhat from the GOE distribution;  however, the pole region presents a slight deviation from the Poisson distribution, similarly as observed for the Pacific pole system. Hence, as happened with the Pacific ocean, the south and central Indian oceans can be understood as set of dynamic subsystems with a strong interaction, while the pole region with slight interaction. }

\section{Discussion and Conclusions}

Analyzing the Equatorial Pacific Ocean, an intrinsic area observed in the present work, Santhanam and Patra \cite{PhysRevE.64.016102} concluded that their results agree with the GOE distribution. The results presented here confirms this conclusions in a broader perspective, since  the eigenvalue statistics of the SST variable in the north, central and southern regions of the Pacific, Atlantic and Indian Oceans are modeled by the universal GOE class of the RMT. We not observe an universality in the spectra fluctuation of nearest-neighbor spacing of SST dynamics, since our results show that the eigenvalue statistics of the SST variable of pole oceans are modeled by the Poisson distribution. This indicates a clear {\it statistical change} of RMT-to-Poisson fluctuations, which was not previously noticed in atmospheric variables.

\begin{figure}
\centering
\includegraphics[width=0.7\linewidth]{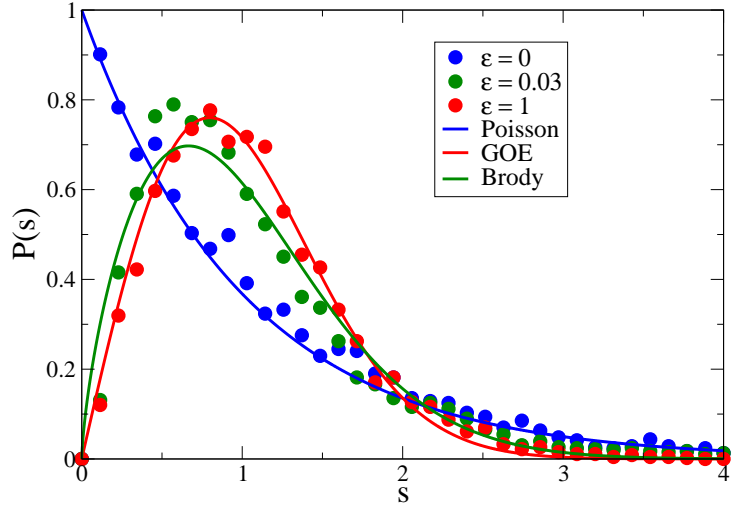}
\caption{ The figure shows the nearest-neighbor spacing histogram of $s_{100}$ for the model of equation (\ref{X}). For $\epsilon = 0$ the histogram is given by Poisson distribution, for $\epsilon = 1$ GOE and for $\epsilon = 0.03$ is a intermediate between RMT-to-Poisson. } \label{fig5}
\end{figure}

To understand a possible cause of {\it statistical change} of  RMT-to-Poisson fluctuations observed in the present work, we developed a model based on reference \cite{PhysRevLett.116.054101}. Let's consider the matrix $X$ of equation (\ref{C}) as a tensor product of two random statistically independent matrices ($X_1$ and $X_2$), where the elements have a gaussian distribution, as following
\begin{equation}
     X(\epsilon)=\left(X_1 \otimes X_2 \right) X_{12}(\epsilon).\label{X}
\end{equation}
The matrix $X_{12}$ is the coupling matrix defined as $X_{12}= \mathbf{1} (1-\epsilon) + \epsilon U$. The $\mathbf{1}$ is identity matrix, $U$ is a random matrix with elements given by a uniform distribution on the interval $\left(-1/2,1/2\right]$ and $\epsilon$ is the intensity of coupling between the two systems that range from 0 to 1. Applying the equation (\ref{X}) in the equation (\ref{C}), we obtain the correlation matrix and its eigenvalues as described above. Figure (\ref{fig5}) presents the nearest-neighbor spacing histogram of $s_{100}$ for different values of interaction ($\epsilon$). Matrices $X_1$ and $X_2$ with dimension $15\times 15$ were used in the numeric simulations, which means that $X$ and $X_{12}$ have dimension $225 \times 225$ and an ensemble with 4,000 matrices. When $\epsilon = 0$ there is no interaction between two systems, hence the fluctuation spacing goes to Poisson distribution. In turn, a strong interaction gives rise to RMT flutuaction (GOE) when $\epsilon = 1$. Furthermore, for {\it slight interaction} $\epsilon = 0.03$ we have a intermediate fluctuation between RMT-to-Poisson.

Following this model, we can understand the SST dynamic as a set of dynamical subsystems with weak or strong interaction. In these RMT regions, the SST is highly correlated and the dynamic subsystems present strong interaction. Therefore, specific $\beta$ values can be used to classify each ocean according to its correlation behavior.

More precisely, the northern boundary matrices for the Pacific and Atlantic oceans presented similar behavior. The north oceanic regions are more geographically closed and they are influenced by the cold waters of the Arctic pole. Therefore, it is possible that these characteristics bring certain randomness in these systems. Randomness is important to determine the level of interaction of eigenvalues in these complex systems, consequently, the correlation among these values.

The oceanographic systems with higher $\beta$ are represented by the regions bounded to the south. The Pacific Ocean stands out for presenting more randomness among the others, and better fit for the universal GOE class described by the Brody distribution.  These regions are more opened and away from continental masses. Thus, they are more susceptible to variations in temperature values.

The values of $\beta$ for the central area of the oceans indicate that they tend to have a less pronounced temporal dynamics compared to the southern region. This was also observed by \citep{ScientificReports.8.14624} studying the globaly SST variation based on a time series model. It seems that the  important and highly researched events that occur in the central areas of the Pacific and Atlantic Oceans, such as El Niño and La Niña, the Atlantic intertropical zone (strictly related to surface temperature of the waters) contribute more to the randomness of their adjacency.

{\it In contrast, the Pacific and Indian pole regions have $\beta$ average significantly lower than other regions ($\beta = 0.43$ and 0.36, respectively), moving away from the RMT fluctuation but without arriving to the Poisson fluctuation. In turn, the Atlantic pole region presented the lowest observed values for the $\beta$ average ($\beta = 0.05$), adjusting well to the Poisson distribution. In these cases, the SST dynamic is composed by a set of dynamical subsystems with slight (Pacific and Indian polo) and weak (Atlantic polo) interactions (independent regions) with more local than global perturbations.}

This seems to be the first observation of such {\it statistical change of RMT-to-Poisson fluctuation} in climate or atmospheric phenomena, such as SST variation. Their correlation matrix spectra can display both Poisson spacing distribution and RMT spacing distribution, formed by subsystems with or without interaction. The consequences of this to climate models need to be investigated.

\section*{Acknowledgements}

This  work  was  partially  supported  by  Brazilian  agencies Conselho  Nacional  de  Desenvolvimento  Cient\'ifico  e  Tecnol\'ogico (CNPq) and supported by  Coordenação de Aperfei\c coamento de Pessoal de N\'ivel Superior (CAPES/DINTER - PROGRAMAS DE DOUTORADO INTERINSTITUCIONAL, Project 20130064: DINTER - UFS - BIOMETRIA E ESTATISTICA APLICADA). 

\section*{References} 
\bibliography{ref}

\begin{thebibliography}{10}
\expandafter\ifx\csname url\endcsname\relax
  \def\url#1{\texttt{#1}}\fi
\expandafter\ifx\csname urlprefix\endcsname\relax\def\urlprefix{URL }\fi
\expandafter\ifx\csname href\endcsname\relax
  \def\href#1#2{#2} \def\path#1{#1}\fi

\bibitem{Biggetal}
G.~R. Bigg, T.~D. Jickells, P.~S. Liss, T.~J. Osborn,
  \href{https://rmets.onlinelibrary.wiley.com/doi/abs/10.1002/joc.926}{The role
  of the oceans in climate}, International Journal of Climatology 23~(10)
  (2003) 1127--1159.
\newblock \href
  {http://arxiv.org/abs/https://rmets.onlinelibrary.wiley.com/doi/pdf/10.1002/joc.926}
  {\path{arXiv:https://rmets.onlinelibrary.wiley.com/doi/pdf/10.1002/joc.926}},
  \href {https://doi.org/10.1002/joc.926} {\path{doi:10.1002/joc.926}}.
\newline\urlprefix\url{https://rmets.onlinelibrary.wiley.com/doi/abs/10.1002/joc.926}

\bibitem{Swarnali}
S.~Majumder, P.~P. Kanjilal,
  \href{https://doi.org/10.1007/s00024-019-02140-4}{Application of singular
  spectrum analysis for investigating chaos in sea surface temperature}, Pure
  and Applied Geophysics 176 (2019) 3769.
\newblock \href {https://doi.org/10.1007/s00024-019-02140-4}
  {\path{doi:10.1007/s00024-019-02140-4}}.
\newline\urlprefix\url{https://doi.org/10.1007/s00024-019-02140-4}

\bibitem{LI201314}
Z.-L. Li, B.-H. Tang, H.~Wu, H.~Ren, G.~Yan, Z.~Wan, I.~F. Trigo, J.~A.
  Sobrino,
  \href{http://www.sciencedirect.com/science/article/pii/S0034425712004749}{Satellite-derived
  land surface temperature: Current status and perspectives}, Remote Sensing of
  Environment 131 (2013) 14 -- 37.
\newblock \href {https://doi.org/https://doi.org/10.1016/j.rse.2012.12.008}
  {\path{doi:https://doi.org/10.1016/j.rse.2012.12.008}}.
\newline\urlprefix\url{http://www.sciencedirect.com/science/article/pii/S0034425712004749}

\bibitem{NASA}
{NASA} national aeronautics and space administration,
  \url{http://www.noaa.gov/}, accessed: 2018-01-28.

\bibitem{Preisendorfer}
R.~W. Preisendorfer, in Principal Component Analysis in Meteorology and
  Oceanography, edited by C.D. Mobley Elsevier, New York, 1988; Daniel S.
  Wilks, Statistical Methods in Atmospheric Sciences Academic Press, London,
  1995.

\bibitem{PhysRevE.64.016102}
M.~S. Santhanam, P.~K. Patra,
  \href{https://link.aps.org/doi/10.1103/PhysRevE.64.016102}{Statistics of
  atmospheric correlations}, Phys. Rev. E 64 (2001) 016102.
\newblock \href {https://doi.org/10.1103/PhysRevE.64.016102}
  {\path{doi:10.1103/PhysRevE.64.016102}}.
\newline\urlprefix\url{https://link.aps.org/doi/10.1103/PhysRevE.64.016102}

\bibitem{Mehta}
M.~L. Mehta, Random Matrices, Elsevier, 2004.

\bibitem{RevModPhys.82.2845}
G.~E. Mitchell, A.~Richter, H.~A. Weidenm\"uller,
  \href{https://link.aps.org/doi/10.1103/RevModPhys.82.2845}{Random matrices
  and chaos in nuclear physics: Nuclear reactions}, Rev. Mod. Phys. 82 (2010)
  2845--2901.
\newblock \href {https://doi.org/10.1103/RevModPhys.82.2845}
  {\path{doi:10.1103/RevModPhys.82.2845}}.
\newline\urlprefix\url{https://link.aps.org/doi/10.1103/RevModPhys.82.2845}

\bibitem{10.2307/1970079}
E.~P. Wigner, \href{http://www.jstor.org/stable/1970079}{Characteristic vectors
  of bordered matrices with infinite dimensions}, Annals of Mathematics 62~(3)
  (1955) 548--564.
\newline\urlprefix\url{http://www.jstor.org/stable/1970079}

\bibitem{PhysRevLett.83.1471}
V.~Plerou, P.~Gopikrishnan, B.~Rosenow, L.~A. Nunes~Amaral, H.~E. Stanley,
  \href{https://link.aps.org/doi/10.1103/PhysRevLett.83.1471}{Universal and
  nonuniversal properties of cross correlations in financial time series},
  Phys. Rev. Lett. 83 (1999) 1471--1474.
\newblock \href {https://doi.org/10.1103/PhysRevLett.83.1471}
  {\path{doi:10.1103/PhysRevLett.83.1471}}.
\newline\urlprefix\url{https://link.aps.org/doi/10.1103/PhysRevLett.83.1471}

\bibitem{PLEROU2000374}
V.~Plerou, P.~Gopikrishnan, B.~Rosenow, L.~Amaral, H.~Stanley,
  \href{http://www.sciencedirect.com/science/article/pii/S0378437100003769}{A
  random matrix theory approach to financial cross-correlations}, Physica A:
  Statistical Mechanics and its Applications 287~(3) (2000) 374 -- 382.
\newblock \href {https://doi.org/https://doi.org/10.1016/S0378-4371(00)00376-9}
  {\path{doi:https://doi.org/10.1016/S0378-4371(00)00376-9}}.
\newline\urlprefix\url{http://www.sciencedirect.com/science/article/pii/S0378437100003769}

\bibitem{PLEROU2001175}
V.~Plerou, P.~Gopikrishnan, B.~Rosenow, L.~Amaral, H.~Stanley,
  \href{http://www.sciencedirect.com/science/article/pii/S037843710100293X}{Collective
  behavior of stock price movements—a random matrix theory approach}, Physica
  A: Statistical Mechanics and its Applications 299~(1) (2001) 175 -- 180,
  application of Physics in Economic Modelling.
\newblock \href {https://doi.org/https://doi.org/10.1016/S0378-4371(01)00293-X}
  {\path{doi:https://doi.org/10.1016/S0378-4371(01)00293-X}}.
\newline\urlprefix\url{http://www.sciencedirect.com/science/article/pii/S037843710100293X}

\bibitem{PhysRevE.65.066126}
V.~Plerou, P.~Gopikrishnan, B.~Rosenow, L.~A.~N. Amaral, T.~Guhr, H.~E.
  Stanley, \href{https://link.aps.org/doi/10.1103/PhysRevE.65.066126}{Random
  matrix approach to cross correlations in financial data}, Phys. Rev. E 65
  (2002) 066126.
\newblock \href {https://doi.org/10.1103/PhysRevE.65.066126}
  {\path{doi:10.1103/PhysRevE.65.066126}}.
\newline\urlprefix\url{https://link.aps.org/doi/10.1103/PhysRevE.65.066126}

\bibitem{STOSIC2018499}
D.~Stosic, D.~Stosic, T.~B. Ludermir, T.~Stosic,
  \href{http://www.sciencedirect.com/science/article/pii/S0378437118305946}{Collective
  behavior of cryptocurrency price changes}, Physica A: Statistical Mechanics
  and its Applications 507 (2018) 499 -- 509.
\newblock \href {https://doi.org/https://doi.org/10.1016/j.physa.2018.05.050}
  {\path{doi:https://doi.org/10.1016/j.physa.2018.05.050}}.
\newline\urlprefix\url{http://www.sciencedirect.com/science/article/pii/S0378437118305946}

\bibitem{PhysRevLett.91.198104}
P.~\ifmmode~\check{S}\else \v{S}\fi{}eba,
  \href{https://link.aps.org/doi/10.1103/PhysRevLett.91.198104}{Random matrix
  analysis of human eeg data}, Phys. Rev. Lett. 91 (2003) 198104.
\newblock \href {https://doi.org/10.1103/PhysRevLett.91.198104}
  {\path{doi:10.1103/PhysRevLett.91.198104}}.
\newline\urlprefix\url{https://link.aps.org/doi/10.1103/PhysRevLett.91.198104}

\bibitem{LUO2006420}
F.~Luo, J.~Zhong, Y.~Yang, R.~H. Scheuermann, J.~Zhou,
  \href{http://www.sciencedirect.com/science/article/pii/S0375960106006530}{Application
  of random matrix theory to biological networks}, Physics Letters A 357~(6)
  (2006) 420 -- 423.
\newblock \href
  {https://doi.org/https://doi.org/10.1016/j.physleta.2006.04.076}
  {\path{doi:https://doi.org/10.1016/j.physleta.2006.04.076}}.
\newline\urlprefix\url{http://www.sciencedirect.com/science/article/pii/S0375960106006530}

\bibitem{PhysRevLett.103.268101}
R.~Potestio, F.~Caccioli, P.~Vivo,
  \href{https://link.aps.org/doi/10.1103/PhysRevLett.103.268101}{Random matrix
  approach to collective behavior and bulk universality in protein dynamics},
  Phys. Rev. Lett. 103 (2009) 268101.
\newblock \href {https://doi.org/10.1103/PhysRevLett.103.268101}
  {\path{doi:10.1103/PhysRevLett.103.268101}}.
\newline\urlprefix\url{https://link.aps.org/doi/10.1103/PhysRevLett.103.268101}

\bibitem{AGRAWAL2014359}
A.~Agrawal, C.~Sarkar, S.~K. Dwivedi, N.~Dhasmana, S.~Jalan,
  \href{http://www.sciencedirect.com/science/article/pii/S037843711301128X}{Quantifying
  randomness in protein–protein interaction networks of different species: A
  random matrix approach}, Physica A: Statistical Mechanics and its
  Applications 404 (2014) 359 -- 367.
\newblock \href {https://doi.org/https://doi.org/10.1016/j.physa.2013.12.005}
  {\path{doi:https://doi.org/10.1016/j.physa.2013.12.005}}.
\newline\urlprefix\url{http://www.sciencedirect.com/science/article/pii/S037843711301128X}

\bibitem{PhysRevE.76.026109}
J.~N. Bandyopadhyay, S.~Jalan,
  \href{https://link.aps.org/doi/10.1103/PhysRevE.76.026109}{Universality in
  complex networks: Random matrix analysis}, Phys. Rev. E 76 (2007) 026109.
\newblock \href {https://doi.org/10.1103/PhysRevE.76.026109}
  {\path{doi:10.1103/PhysRevE.76.026109}}.
\newline\urlprefix\url{https://link.aps.org/doi/10.1103/PhysRevE.76.026109}

\bibitem{PhysRevE.96.030101}
A.~Jagannath, T.~Trogdon,
  \href{https://link.aps.org/doi/10.1103/PhysRevE.96.030101}{Random matrices
  and the new york city subway system}, Phys. Rev. E 96 (2017) 030101.
\newblock \href {https://doi.org/10.1103/PhysRevE.96.030101}
  {\path{doi:10.1103/PhysRevE.96.030101}}.
\newline\urlprefix\url{https://link.aps.org/doi/10.1103/PhysRevE.96.030101}

\bibitem{GONZALEZ20172912}
R.~E. González, I.~A. Santos, M.~G. Nunes, V.~M. de~Oliveira, A.~L. Barbosa,
  \href{http://www.sciencedirect.com/science/article/pii/S0375960117306874}{Statistical
  behavior of time dynamics evolution of hiv infection}, Physics Letters A
  381~(35) (2017) 2912 -- 2916.
\newblock \href
  {https://doi.org/https://doi.org/10.1016/j.physleta.2017.07.022}
  {\path{doi:https://doi.org/10.1016/j.physleta.2017.07.022}}.
\newline\urlprefix\url{http://www.sciencedirect.com/science/article/pii/S0375960117306874}

\bibitem{CHATTERJEE20181352}
S.~Chatterjee, P.~Barat, I.~Mukherjee,
  \href{http://www.sciencedirect.com/science/article/pii/S0378437117311482}{Universality
  in the dynamical properties of seismic vibrations}, Physica A: Statistical
  Mechanics and its Applications 492 (2018) 1352 -- 1363.
\newblock \href {https://doi.org/https://doi.org/10.1016/j.physa.2017.11.062}
  {\path{doi:https://doi.org/10.1016/j.physa.2017.11.062}}.
\newline\urlprefix\url{http://www.sciencedirect.com/science/article/pii/S0378437117311482}

\bibitem{PhysRevLett.116.054101}
S.~C.~L. Srivastava, S.~Tomsovic, A.~Lakshminarayan, R.~Ketzmerick,
  A.~B\"acker,
  \href{https://link.aps.org/doi/10.1103/PhysRevLett.116.054101}{Universal
  scaling of spectral fluctuation transitions for interacting chaotic systems},
  Phys. Rev. Lett. 116 (2016) 054101.
\newblock \href {https://doi.org/10.1103/PhysRevLett.116.054101}
  {\path{doi:10.1103/PhysRevLett.116.054101}}.
\newline\urlprefix\url{https://link.aps.org/doi/10.1103/PhysRevLett.116.054101}

\bibitem{CHATTERJEE2019122189}
S.~Chatterjee, I.~Mukherjee,
  \href{http://www.sciencedirect.com/science/article/pii/S0378437119312683}{Manifestation
  of crossover from rmt fluctuation to poisson fluctuation in bse sensex, a
  prototype of financial systems}, Physica A: Statistical Mechanics and its
  Applications 534 (2019) 122189.
\newblock \href {https://doi.org/https://doi.org/10.1016/j.physa.2019.122189}
  {\path{doi:https://doi.org/10.1016/j.physa.2019.122189}}.
\newline\urlprefix\url{http://www.sciencedirect.com/science/article/pii/S0378437119312683}

\bibitem{NOAA}
{NOAA} national oceanic and atmospheric administration,
  \url{http://www.nasa.gov/}, accessed: 2017-01-27.

\bibitem{doi:10.1175/2007JCLI1824.1}
R.~W. Reynolds, T.~M. Smith, C.~Liu, D.~B. Chelton, K.~S. Casey, M.~G. Schlax,
  \href{https://doi.org/10.1175/2007JCLI1824.1}{Daily high-resolution-blended
  analyses for sea surface temperature}, Journal of Climate 20~(22) (2007)
  5473--5496.
\newblock \href {http://arxiv.org/abs/https://doi.org/10.1175/2007JCLI1824.1}
  {\path{arXiv:https://doi.org/10.1175/2007JCLI1824.1}}, \href
  {https://doi.org/10.1175/2007JCLI1824.1} {\path{doi:10.1175/2007JCLI1824.1}}.
\newline\urlprefix\url{https://doi.org/10.1175/2007JCLI1824.1}

\bibitem{Brody1973}
T.~A. Brody, \href{https://doi.org/10.1007/BF02727859}{A statistical measure
  for the repulsion of energy levels}, Lettere al Nuovo Cimento (1971-1985)
  7~(12) (1973) 482--484.
\newblock \href {https://doi.org/10.1007/BF02727859}
  {\path{doi:10.1007/BF02727859}}.
\newline\urlprefix\url{https://doi.org/10.1007/BF02727859}

\bibitem{PhysRevE.72.045204}
J.~Sakhr, J.~M. Nieminen,
  \href{https://link.aps.org/doi/10.1103/PhysRevE.72.045204}{Poisson-to-wigner
  crossover transition in the nearest-neighbor statistics of random points on
  fractals}, Phys. Rev. E 72 (2005) 045204.
\newblock \href {https://doi.org/10.1103/PhysRevE.72.045204}
  {\path{doi:10.1103/PhysRevE.72.045204}}.
\newline\urlprefix\url{https://link.aps.org/doi/10.1103/PhysRevE.72.045204}

\bibitem{ScientificReports.8.14624}
P.~K. Dunstan, S.~D. Foster, E.~King, J.~Risbey, T.~J. O’Kane, D.~Monselesan,
  A.~J. Hobday, J.~R. Hartog, P.~A. Thompson,
  \href{https://doi.org/10.1038/s41598-018-33057-y}{Global patterns of change
  and variation in sea surface temperature and chlorophyll a}, Sci. Rep. 8
  (2018) 14624.
\newblock \href {https://doi.org/10.1038/s41598-018-33057-y}
  {\path{doi:10.1038/s41598-018-33057-y}}.
\newline\urlprefix\url{https://doi.org/10.1038/s41598-018-33057-y}

\end{thebibliography}

\end{document}